\documentclass[aps,prphys,reprint,superscriptaddress]{revtex4-2}
\usepackage{graphicx}
\usepackage{amsmath}
\usepackage{amssymb}
\usepackage{siunitx}
\usepackage[dvipsnames]{xcolor}
\usepackage[version=3]{mhchem} %

\usepackage[hidelinks=true]{hyperref}
\usepackage{mathpazo}  %
\usepackage{microtype}

\title{Unveiling and quantifying the topology-dependent premelting of nanoparticles}
               

\definecolor{mycolor}{HTML}{e64553}

\begin{document}
    \author{Marthe Bideault}
    \affiliation{%
    Materials Design SARL, 42 avenue Verdier, 92120 Montrouge, France
    }%
    \affiliation{%
    ICMMO, Université Paris-Saclay, UMR 8182, 17 avenue des Sciences, 91400 Orsay, France
    }%
    \author{Arnaud Allera}
    \affiliation{%
    ASNR/PSN-RES/SEMIA/LSMA Centre d’études de Cadarache, F-13115 Saint Paul-lez-Durance, France
    }
    \email{arnaud.allera@asnr.fr}
    \author{Ryoji Asahi}
    \affiliation{%
    Institute of Materials Innovation, Nagoya University, Nagoya 464-8603, Japan
    }%
    \author{Jérôme Creuze}
    \affiliation{%
    ICMMO, Université Paris-Saclay, UMR 8182, 17 avenue des Sciences, 91400 Orsay, France
    }%
    \email{jerome.creuze@universite-paris-saclay.fr}
    \author{Erich Wimmer}
    \affiliation{%
    Materials Design SARL, 42 avenue Verdier, 92120 Montrouge, France
    }%
    \email{ewimmer@materialsdesign.com}
    \affiliation{%
    Materials Design, Inc., 12121 Scripps Summit Drive, \#160, San Diego, California 92131, USA }%

\begin{abstract}
    The melting of metallic nanoparticles is governed by surface premelting, a phenomenon traditionally modeled as the isotropic growth of a uniform liquid shell. 
    Challenging this classical view, we report facet-dependent premelting in hexagonal close-packed Co nanoparticles, arising from the structural heterogeneity of their surface.
    In molecular dynamics simulations (587 to 13047 atoms), the onset of surface mobility is observed as low as 20\% of the bulk melting point, driven by the early disordering of stepped $\{01\bar{1}1\}$ facets.
    These facets consistently melt nearly 150~K below flat $\{0001\}$ facets, regardless of particle size.
    We show that both surface and facet melting temperatures scale with nanoparticle size through the Gibbs--Thomson effect, and determine a size-dependent critical liquid layer thickness that triggers complete melting of the nanoparticle, which saturates near three atomic layers.
    Our results confirm recent experimental observations of surface premelting and extend the framework to anisotropic particles with facet-orientation-dependent behavior.

\end{abstract}
\title{Unveiling and quantifying the topology-dependent premelting of nanoparticles}

\maketitle

\section{\label{sec:intro}Introduction}

Nanoparticles are essential components across a wide range of fields, including alloys, catalysis, electronics, optics, and magnetic materials, as well as medical and cosmetic applications. Their high surface-to-volume ratio imparts unique size-dependent properties, such as tunable optical and magnetic responses \cite{magnetic-np}, increased hardness \cite{hardness}, and melting point depression \cite{mp-depression, mpnp-exp}. The latter is described by the classical Gibbs--Thomson relation \cite{mpnp-exp, CoP}, which predicts that the melting point of a spherical nanoparticle scales as $N^{-1/3}$, where $N$ is the number of atoms.

The melting behavior of nanoparticles has been extensively studied, beginning with the pioneering work of Pawlow in 1909, who assumed a global transformation characterized by a single transition temperature \cite{Pawlow}. 
In 1924, Tammann and Hüttig were the first to suggest that melting was a surface phenomenon \cite{tammann}. They introduced approximate temperatures at which bulk (Tammann) and surface (Hüttig) atoms begin to diffuse in a solid material with melting temperature $T_{m,\infty}$, expressed as $T_{\text{Tammann}} = 0.5 \times T_{m,\infty}$ and $T_{\text{Hüttig}} = 0.3\times T_{m,\infty}$, respectively.
Later, in 1948, Reiss and Wilson proposed a model in which a thin liquid layer forms on the surface of the nanoparticle at lower temperatures and persists until complete melting \cite{Reiss-Wilson}. 
In 1977, Couchman and Jesser introduced a model in which this liquid layer grows inward toward the core \cite{Couchman-Jesser}. Subsequent experimental and theoretical studies have confirmed that nanoparticle melting is governed by surface premelting \cite{mpnp-exp, mp-depression, np-melting-1-surf, np-melting-2-surf, np-melting-3-surf}.

In surface science, it is well-established that surfaces can exhibit complete premelting \cite{full-pre-melt}, incomplete premelting \cite{semi-pre-melt}, or even no premelting \cite{non-pre-melt}. However, when modeling nanoparticles, this behavior is comparatively less explored: surface melting is often considered as a global phenomenon, treating the entire nanoparticle surface as one entity \cite{baletto, np-melting-2-surf, np-melting-3-surf, unique-surf-1}, although surfaces can consist of multiple distinct facet types. As a result, certain facets may melt at lower temperatures, leading to surface reconstructions or phenomena of interest for catalysis, where nanoparticles operate at relatively high temperatures. 

Electronic microscopy imaging recently showed direct evidence of the premelting of Sn nanoparticles~\cite{sn-np-melting}, where a quasi-liquid layer forms at the surface and remains stable until full melting near \SI{500}{\kelvin}. 
The effect of nanoparticle size and surface orientation remains difficult to resolve from these approaches, and similar experiments are currently lacking for metallic nanoparticles melting at higher temperature, such as Co, motivating numerical modeling. 

In this study, hexagonal close-packed (hcp) cobalt nanoparticles are used as prototypical systems. They exhibit two different types of facets: stepped $\{01\bar{1}1\}$ facets and flat $\{0001\}$ facets. 
The corresponding surface energies are 2.13 and 2.38 J.m\textsuperscript{-2} respectively, according to computations with a q-SNAP machine-learned interatomic potential introduced recently \cite{CoP}, which was trained on density functional theory (DFT) calculations using the Perdew-Burke-Ernzerhof (PBE) functional \cite{pbe}. The surface energies computed directly with DFT-PBE yield nearly the same surface energies, namely 2.13 and 2.40 J.m\textsuperscript{-2}, respectively. For comparison, the face-centered cubic (fcc) (111) surface of cobalt, which is the most stable surface for this metal, has a slightly lower surface energy of 2.07 (DFT-PBE) or 2.09 (q-SNAP) J.m\textsuperscript{-2}. However, below 700~K, cobalt favors the hexagonal lattice structure, making hcp nanoparticles stable at low temperatures. This hexagonal shape is observed to be stable for relatively large cobalt nanoparticles, both experimentally \cite{co-hcp-np} and computationally \cite{farkas, CoP}, thus justifying the use of hcp NPs throughout the present study.

Isolated nanoparticles can be simulated at the atomic scale using molecular dynamics (MD) \cite{baletto} (see Methods). 
The main challenge for accurate MD simulations is the realism of the interatomic potential used to describe interactions between atoms. This bottleneck was recently lifted by the introduction of an accurate machine-learned interatomic potential dedicated to Co, which closely matches DFT-PBE calculations over a range of properties, including surface energies, phonon dispersions, phase transition temperature, melting point, vacancy formation energies and elastic coefficients \cite{CoP}. Another important challenge for atomistic studies, and the primary focus of this study, is the analysis of simulation data, which requires the identification of atomic patterns representative of crystal structures or surface orientations.

To characterize crystals, algorithms such as polyhedral template matching (PTM) \cite{ptm}, adaptive common neighbor analysis (a-CNA) \cite{a-cna}, or bond order parameters \cite{bond-order-param,bop-shorcomings}, are commonly used, to provide a local and computationally efficient geometric analysis of bulk materials. 
These methods have been applied to the structural characterization of the surface and the melting of metallic NPs \cite{surf-struct-ana}, along with the analysis of the radial distribution function \cite{rdf-ana}, the surface coordination number \cite{scn-ana,unique-surf-1} and/or the bond length fluctuation index (also called Lindemann index) \cite{model-morpho-np,unique-surf-1}. The latter is a traditional measure of interatomic distance fluctuations over the neighborhood of each atom in the NP, used to identify transitions from order to disorder states, such as crystalline and liquid ones. A departure from linearity of the global Lindemann index, i.e., averaged over all atoms, or an increase above an \textit{ad hoc} threshold is taken as the signature of a solid-liquid phase transition \cite{unique-surf-1}. 

Recently, more sophisticated methods associating machine learning with atomic descriptors such as a-CNA \cite{roncaglia2023machine}, radial distribution functions \cite{telari2023charting} or smooth overlap of atomic positions (SOAP) \cite{soap,soapia-gold-surf,soapia-gold-np} have been developed to overcome the major drawbacks of the previously cited standard approaches.
Some of these approaches remain too sensitive to thermal fluctuations, crystal defects and large deformations appearing in NPs to be applied on surface melting \cite{a-cna, thermal-noise}. An approach based on atomic cluster expansion (ACE) descriptors and a hierarchical k-means clustering achieved improved temperature stability, allowing structural analysis up to the melting point of gold NPs \cite{soapia-gold-melting}.

Modern structural analysis relies on high dimensional invariant representations of local atomic environments \cite{soap,bso4,soapia-gold-surf,soapia-gold-melting} or atomic descriptors, which enable various machine-learning strategies. In particular, atomic descriptors have demonstrated strong performance for atomic structure recognition, owing to their flexibility and improved robustness against thermal fluctuations \cite{neighbor-map, robust, distortion-score, tom}. Descriptors such as the BSO(4) bispectral descriptor functions \cite{bso4}, encode atomic neighborhoods as compact feature vectors $\mathbf{v} \in \mathbb{R}^d$, typically $d=55$ for BSO(4) descriptors using $j_{max}=4$, which remains tractable without necessarily requiring dimensionality reduction techniques. 
Atomic descriptors can be robustly compared to distributions of interest following Gorayeva \emph{et. al.} using distortion scores \cite{distortion-score} based on the analysis of Mahalanobis distances (see the Methods section). This approach allows the characterization of e.g., crystal defects in solids by measuring how much an atomic environment deviates from predefined statistical distributions, while being robust to fluctuations. 
This method, primarily aimed at the detection and analysis of structural defects in crystals, rests on the assumption of unimodal multivariate Gaussians. It has yet to be extended to more complex systems, where there is a coexistence of many distributions with various population sizes.

In this study, we use a hierarchical approach based on the distortion score method to analyze the inhomogeneous melting of hcp cobalt nanoparticles, ranging from 587 to 13,047 atoms. 
Qualitatively, we observe that atoms of the $\{01\bar{1}1\}$ facets become mobile at low temperature, with vertex atoms beginning to diffuse on top of these facets around 400 K. 
In contrast, the denser $\{0001\}$ facets remain stable and organized up to significantly higher temperatures .
Another striking feature is the premelting of nanoparticles surface, where a liquid layer is formed at the surface, with a thickness that increases linearly with the nanoparticle size up to almost 5000 atoms, and subsequently tends towards a limiting value of approximately three atomic layers for larger nanoparticle sizes.
In view of these phenomena, the goal of this study is to propose a quantitative analysis of the differential surface melting, and to investigate the premelting behavior of Co nanoparticles. 

Our analysis protocol consists of two simple steps and does not rely on external pretraining or manual data labeling. First, we use an unsupervised machine-learning approach to establish reference statistical distributions corresponding to the main atomic environments present at low temperature, including atoms in the core, at $\{01\bar{1}1\}$ and $\{0001\}$ facets, edges, and vertices. 
We then classify atoms during heating simulations, based on these reference atomic structures, using robust distortion scores. 
Atoms that significantly differ from all reference structures (with respect to the Mahalanobis distance) are noted as outliers, avoiding ambiguities in labeling non-crystalline environments.
Our approach is easily extensible and can be easily applied to other systems presenting multiple distributions of atomic patterns. 

The subsequent sections are organized as follows: we begin by exploring the global melting behavior of hcp cobalt nanoparticles and validating our approach.
Then, we present a detailed analysis of differential surface melting, and surface premelting. 
We show that both effects can be attributed to the Gibbs-Thomson effect.
premelting is further characterized and discussed with respect to existing theories and experimental works.  
We describe our simulation and analysis methodology in the Methods section.

\begin{figure*}[ht!]
    \centering
    \includegraphics[width=0.85\textwidth]{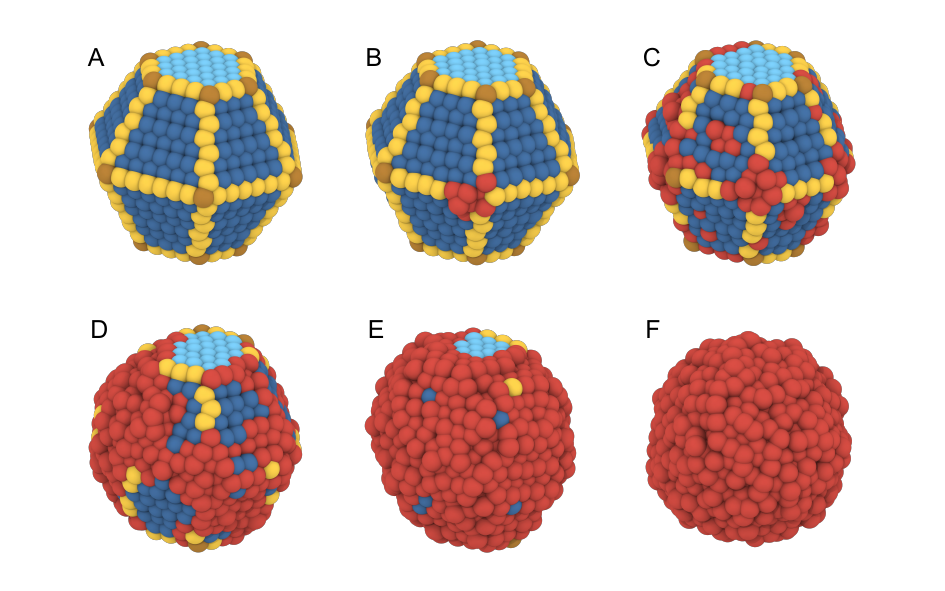}
    \caption{Facet-dependent surface premelting in a 1483 atoms hexagonal close-packed cobalt nanoparticle during a heating simulation. Snapshots are taken at 0 K (A), 370 K (B), 750 K (C), 1000 K (D), 1260 K (E) and 1380 K (F), which is the melting point. Atoms belonging to $\{0001\}$ facet are colored in light blue, those from $\{01\bar{1}1\}$ facet in dark blue, edge atoms are yellow and vertices are brown. Atoms that does not belong to any class are colored in red: they are outliers.} 
    \label{fgr:trj}
\end{figure*}
\section{Results}\label{sec:results}

\subsection{Classification of atoms}
\label{subsec:classification}

In the present work, we seek to develop a classification scheme, where individual atoms are assigned labels by a model, allowing intuitive interpretation, while avoiding the explicit labeling of individual atoms on a training dataset.
We use an unsupervised approach to define reference classes of local atomic environments, which are then tracked during dynamics. 
This task is challenging in nanoparticles due to their complex geometry consisting in a rich variety of local environments. 
In contrast to simple periodic crystals, where symmetrically equivalent atoms typically produce unimodal distributions of atomic descriptors \cite{distortion-score}, the diverse positions on nanoparticle surfaces result in many non-equivalent local environments, even within the same crystallographic facet or edge.

To address this issue, we adopt a hierarchical approach based on Gaussian Mixture Models (GMMs).
While GMMs can in principle be trained to optimize a mixture of $k$ Gaussian distributions, hence describing multimodal distributions, this is difficult in the present case due the strong imbalance of classes populations. 
In the example shown in frame A of Fig.~\ref{fgr:trj}, only 18 atoms belong to the vertex class, while up to 967 atoms belong to the bulk class. 

We take advantage of the naturally hierarchical distribution of nanoparticle environments, where facets, edges, and vertices form a broader class of surface atoms (see frame A of Fig.~\ref{fgr:trj}), to implement an iterative classification scheme suitable for imbalanced multimodal distributions. 
This iterative approach achieves a stratified decomposition of the descriptor space,  using local Gaussians to approximate the multi-scale structure of the data distribution.
The entire hierarchy is illustrated in Fig.~S1.
To investigate complex alloys, Poisvert et al.~\cite{poisvert2025short} addressed multimodality using a $K$-dimensional extension of the Mahalanobis distance, on which GMM clustering was applied. Here, we found a direct hierarchical approach to be sufficient. 

After identifying reference distributions at low temperature, we analyze MD trajectories by comparing local atomic environment descriptors to these reference classes, using a statistical distance criterion. Specifically, we apply a Mahalanobis distance threshold to distinguish in-distribution and out-of-distribution samples \cite{distortion-score}. Environments with large distances to all known classes are flagged as outliers, representing configurations significantly different from low temperature environments. For example, when a vertex atom migrates across the $\{01\bar{1}1\}$ surface, the atom and its close neighbors become classified as outliers (see Fig.~1B). As the nanoparticle melts, the number of outlier atoms increases, eventually encompassing the entire system (see Fig.~1F).

This flexible method allows for a robust analysis of complex, high-temperature behavior, such as diffusion, defects, and local disorder, where rigid labels like \emph{crystalline}, \emph{defective} or \emph{melted} become ambiguous.
However, for simplicity throughout the text, we use the terms melted or liquid region interchangeably to refer to the outlier-rich region that appears at the surface during premelting and eventually extends to the core.
Fig.~\ref{fgr:trj} illustrates this evolution through representative snapshots at increasing temperatures, from 0 to 1380 K.

\subsection{Global melting}
\label{subsec:global_melting}

\begin{figure}
    \centering
    \includegraphics[width=\columnwidth]{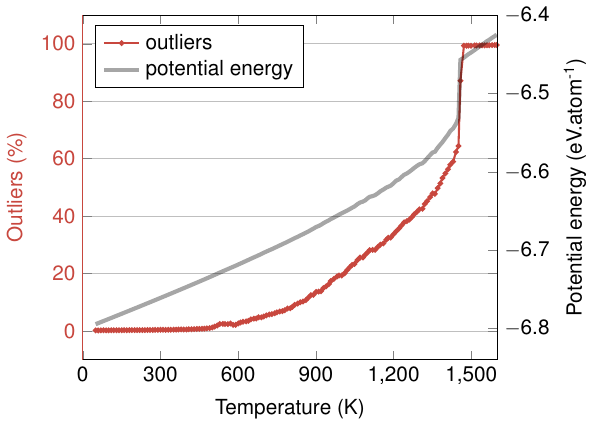}
    \caption{Percentage of outliers (plain red line) as a function of temperature, for a 3009 atoms hcp nanoparticle. 
    The gray dashed line corresponds to the potential energy of the system. 
    }
  \label{fgr:doutliers}
\end{figure}

\begin{figure}
    \centering
    \includegraphics[width=0.9\columnwidth]{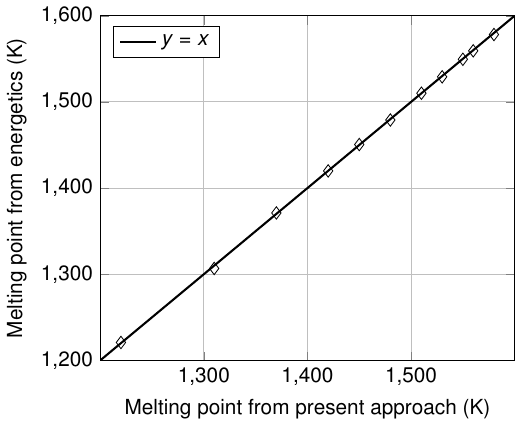}
    \caption{Correlation between melting point from the present structural approach and from energetics, i.e., the maximum of the heat capacity.}
    \label{fgr:melting_p_vs_nrj}
\end{figure}

Upon heating, nanoparticles undergo structural transformations from a crystalline arrangement to a liquid state, with surface premelting playing a critical role. 
In this study, hcp Co nanoparticles were heated from 50 to 1600~K using molecular dynamics simulations (see Methods). Atoms were classified based on the structural distributions established at training temperatures between 50 and 400~K. Atoms recognized as out-of-distribution for all classes were labeled as outlier atoms.
Starting above 400~K, where surface diffusion begins, the fraction of outlier atoms (with respect to the entire nanoparticle) increases, reflecting atomic rearrangements on the surface, as illustrated in Fig.~\ref{fgr:trj}. 
Starting at about 600 K, the percentage of outliers increases with a smoothly increasing slope as temperature rises, before a sudden steep increase signals the complete melting of the nanoparticle. 
The temperature where the fraction of outliers jumps corresponds to the global melting temperature $T_{m,NP}$, as shown in Fig.~\ref{fgr:doutliers}. 
The melting temperature derived using this method is in excellent agreement with that obtained from the maximum of the heat capacity, as demonstrated in Fig.~\ref{fgr:melting_p_vs_nrj}.

In (finite size) nanoparticles, the melting temperature is size-dependent due to the Gibbs-Thomson effect \cite{mpnp-exp,mpnp-simu,de2023melting,rusanov2005surface}, as we succinctly recall below.
The Gibbs energy of a phase $i$ occupying a finite domain results from a core (approximated as bulk) and a surface contribution 
        \begin{equation}
        G_i = G_i^{b} + G_i^{s} =  n_i \mu_i^b + \sigma_i A_i,
        \end{equation}
        where $A_i$ is the surface area, $\sigma_i$ the surface tension, $\mu_i^b$ the bulk chemical potential, and $n_i$ the number of atoms in phase $i$.
        The chemical potential $\mu_i = (\partial G_i/\partial n_i)_{T,P}$ is
        
        \begin{equation}
            \mu_i = \mu^b_i + \sigma_i \left(\frac{\partial A_i}{\partial n_i}\right)_{T,P}.
        \end{equation}

        Assuming a nanoparticle melting from its surface, the melting temperature can be expressed as a function of particle size (see \cite{li2025size} for a full derivation of the model).
        During melting, the change in molar Gibbs energy due to the transformation of solid (written $s$) into liquid ($l$) is
    \begin{align}
        \Delta^{s\rightarrow l} G_m &= \mu_l -\mu_s \\
        &= \Delta^{s\rightarrow l} G_m^{b} + \sigma_l \left(\frac{\partial A_l}{\partial n_l}\right)_{T,P} - \sigma_s \left(\frac{\partial A_s}{\partial n_s}\right)_{T,P},
    \end{align}
        which reflects three contributions: the melting of the bulk solid phase $\Delta^{s\rightarrow l} G_m^{b}$, 
        the increase of the (outer) surface energy associated with the liquid, 
        and the reduction of the surface energy associated with the solid (at the solid-liquid interface).  
        Near the melting point, it is assumed that the density of the liquid and solid phase are equal, and that melting doesn't increase the liquid surface area, i.e. $(\partial A_l / \partial n_l) \approx 0$.

        Assuming equilibrium of the two phases, i.e. $\mu_l = \mu_s$, and writing 
        $G_m^{b} = H_m^{b} - TS_m^{b}$, 
        the characteristic melting temperature can be written:
        \begin{equation}
            T_{m,NP} = T_{m,\infty} - \frac{\sigma_s}{\Delta^{s\rightarrow l} S_m^b}   \left(\frac{\partial A_s}{\partial n_s}\right)_{T,P},
        \end{equation}
        where the bulk characteristic melting temperature $T_{m,\infty}=\Delta^{s\rightarrow l} H_m^b / \Delta^{s\rightarrow l} S_m^b$ is constant.
        
        The equation above applies to a finite nanoparticle melting from its surface, assuming no particular shape.
        For spherical nanoparticles with radius $r$, we decompose
        \begin{equation}
        \left(\frac{\partial A_s}{\partial n_s}\right)_{T,P} = \left(\frac{\partial A_s}{\partial r}\right) _{T,P}\left(\frac{\partial r}{\partial n_s}\right)_{T,P}.
        \end{equation}

\begin{figure}[hbtp]
    \centering
    \includegraphics[width=\columnwidth]{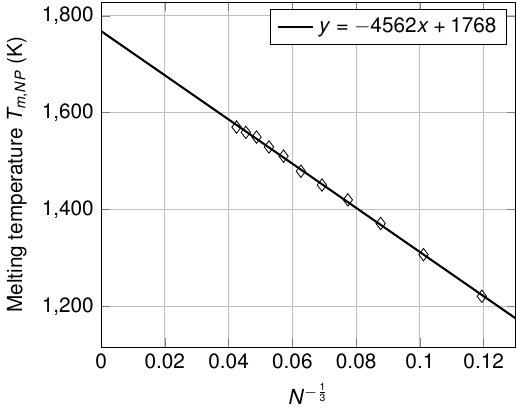}
    \caption{Linear regression of the melting points of nanoparticles as a function of $N^{-1/3}$, where $N$ is the number of atoms. The linear fit intersect the $y$ axis at 1768 K.}
    \label{fgr:GT3D}
\end{figure}

        The surface area is $A_s = 4\pi r^2$ and $n_s = V/v_m = 4/3\pi  r^3 / v_m$, with $v_m$ the atomic volume. 
        It follows that $r = \sqrt[3]{n_s v_m / (4/3\pi)} $. It is direct to show that $(\partial A_s / \partial r) \propto n_s^{1/3} $, and that
        $(\partial r/\partial n_s) 
        \propto n_s^{-2/3}$. %
        As a result, for spherical nanoparticles created with $N\equiv n_s$ atoms, we obtain the usual Gibbs-Thomson form
\begin{equation}
    T_{m,\text{NP}} = T_{m,\infty} \left (1 - aN^{-1/3}\right),
\label{eqn:eq2}
\end{equation}
with $a$ denoting a material-dependent constant.
As shown in Fig.~\ref{fgr:GT3D}, this relation is well satisfied in our simulations, with melting temperatures decreasing linearly with $N^{-1/3}$. The linear fit intersects the $y$-axis at $T = 1768$ K, which can be interpreted as the infinite size limit, i.e., the bulk melting temperature of hcp cobalt.

Nevertheless, bulk cobalt undergoes an allotropic transition from hcp to fcc when heated above 700~K \cite{tdp-700}, which we did not observe in the present study, as evidenced by our structural analysis methodology. 
We hypothesize that the high-temperature stability of hcp for the sizes considered here is favored by the initial hcp geometry of the nanoparticles, which are created with a number of atoms chosen to form a regular shape, and due to the compressive stress of the core induced by the surface in finite nanoparticles.
Actually, the present interatomic potential reproduces well both the experimental bulk melting point of fcc Co and the hcp to fcc transition, as shown in our previous study \cite{CoP}.

\begin{figure}[hbp]
    \includegraphics[width=\columnwidth]{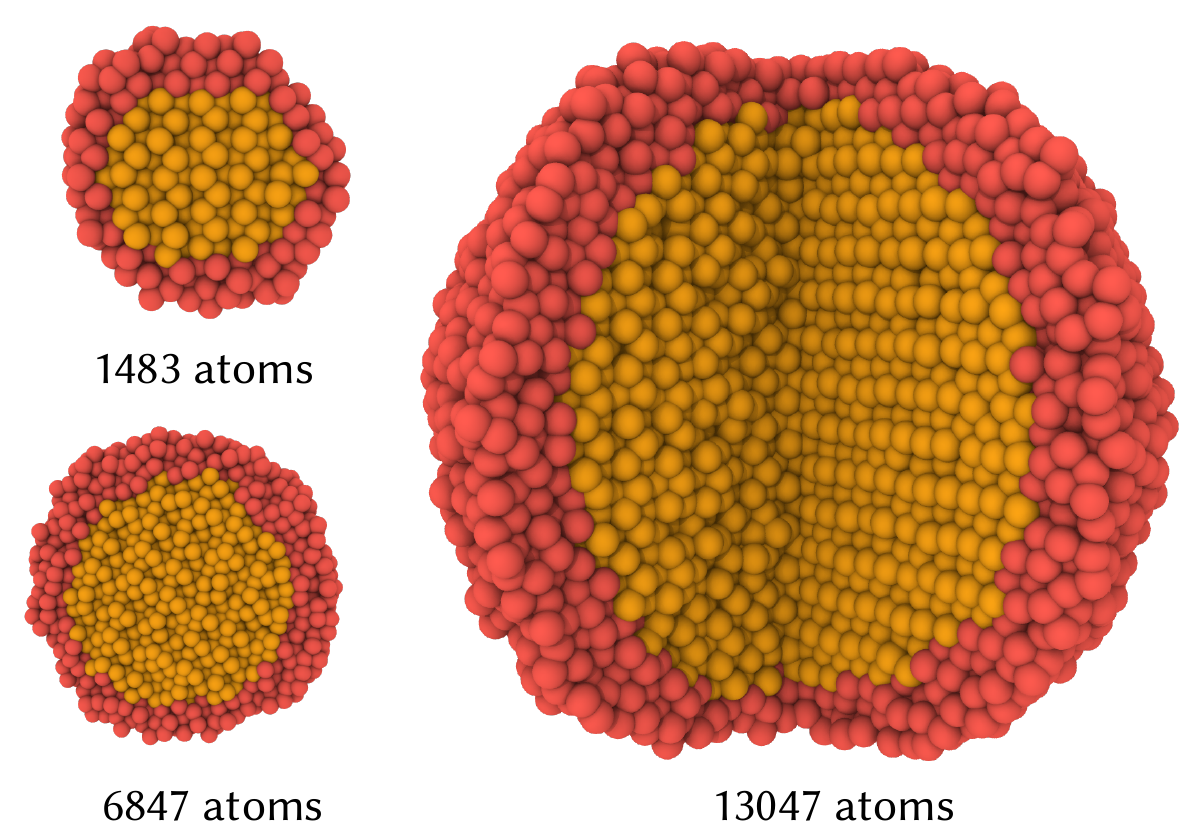}
    \caption{Snapshots taken at the highest temperature before complete melting, for different nanoparticle sizes. Red atoms are atoms classified as outliers in more than 50\% of the trajectory frames. 
    The thickness of the liquid layer (external red layer) increases with nanoparticle size. 
    }
    \label{fgr:just_before_melting}
\end{figure}

\subsection{Surface melting}\label{subsec:surface_melting}

Beyond the global nanoparticle melting, our approach enables the detailed analysis of surface premelting.
In Fig.~\ref{fgr:doutliers}, it can be seen that the fraction of outliers reaches a maximum before a jump associated to melting.
The maximum thickness of the liquid layer at this critical state is found to increase with the nanoparticle size, as illustrated in 
Figs.~\ref{fgr:just_before_melting} and S4.

We use the OVITO \cite{ovito} software to directly extract the liquid layer thickness from our simulations.
For each nanoparticle size, we use the MD trajectory conducted at the highest (constant) temperature below $T_{m,NP}$.
To robustly determine the thickness of the melted zone, we consider the atoms that are classified as outliers in more than 50\% of the trajectory frames as being part of the liquid layer. 
We then create two surface meshes, containing the full nanoparticle or excluding the liquid layer, respectively.
Assuming that the volumes contained by each surface, noted $V_{\mathrm{NP}}$ for the full NP and $V_{\mathrm{core}}$ for the core, have a spherical shape and share the same center, we compute an effective layer thickness for an ideal spherical NP, as the difference between the NP radius and its core radius, $l = R_{\mathrm{NP}}-R_{\mathrm{core}} = (V_{\mathrm{NP}}/(4\pi/3))^{1/3} - (V_{\mathrm{core}}/(4\pi/3))^{1/3}$.

\begin{figure}[b]
    \centering
    \includegraphics[width=0.90\columnwidth]{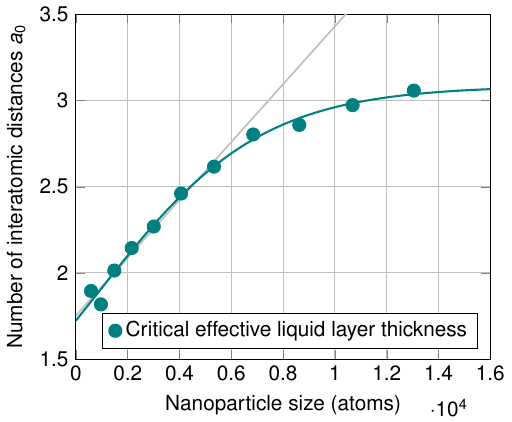}
    \caption{Critical effective liquid layer thickness as a function of nanoparticle size. The circular markers are thickness values extracted using the OVITO software \cite{ovito}, the green solid line is a fit to a sigmoid function  $y = L (1 + \exp(-k (x - x_0)))^{-1} + b$ and the light gray line is a linear fit below 6,000 atoms. Both serve as a guide for the eye to ease interpretation.}
    \label{fig:effective-thickness}
\end{figure}

The dependence of the critical effective liquid layer thickness with NP size is shown in Fig.~\ref{fig:effective-thickness}, normalized by the zero Kelvin lattice parameter, i.e. $a_0 = \SI{2.49}{\angstrom}$ for the present potential \cite{CoP}.
Our analysis shows that the thickness increases monotonously, but that beyond a linear domain that extends up to 5000 atoms, the increase becomes much slower, suggesting a constant thickness for the larger NP sizes.
This unexpected change of regime reflects the unfavorable cost of maintaining a liquid layer when reaching a thickness about three times the lattice parameter.
We note that for the two smallest sizes, the obtained layer is more anisotropic, making the methodology less reliable (see Fig.~S4).

\begin{figure}[b]
    \centering
    \includegraphics[width=0.95\columnwidth]{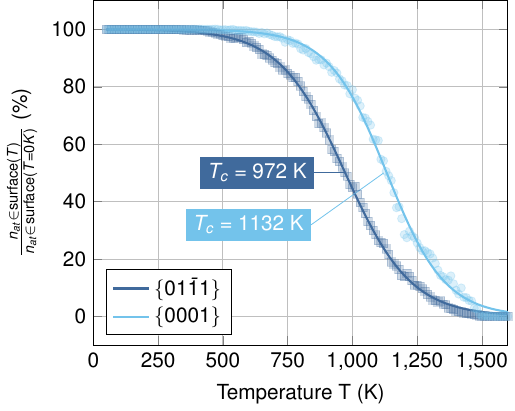}
    \caption{Percentage of atoms that belong to 
    $\{01\bar{1}1\}$ (dark blue) and $\{0001\}$ (light blue) facets. Plain lines correspond to sigmoid fit of these data. Characteristic temperatures $T_{c(hkil)}$ are indicated for each surface.}
    \label{fgr:surface_melting}    
\end{figure}

The analysis of surface premelting can be further refined by considering facet-dependent melting. 
Hcp nanoparticles have two types of facets: flat $\{0001\}$ facets and stepped 
$\{01\bar{1}1\}$ facets, as illustrated in Fig.~\ref{fgr:trj}. The present method enables one to identify the melting of each facet independently. Fig.~\ref{fgr:surface_melting} shows the percentage of atoms that remain within their initial facet as a function of temperature, for a 4061 atoms nanoparticle (see Fig.~S3 for the other sizes). 
For both facets, the evolution of the fraction of atoms exhibits a continuous decrease, with a sigmoid-shaped curve that is translated to higher temperatures for the flat $\{0001\}$ facet.
For each size, we seek to extract a characteristic melting temperature, using the following sigmoid function:
\begin{equation}
  \dfrac{n_{at\in facet}(T)}{n_{at\in facet}(T=0\ \text{K})} = -\frac{L}{1+e^{-k(T-T_c)}}+b, \label{eqn:sigmoid}
\end{equation}
where \(L\), \(k\), \(b\) and \(T_c\) are fitting parameters.
The temperature corresponding to the maximum slope of the sigmoid function is termed the characteristic temperature \(T_c\).

\begin{figure}
    \centering
    \includegraphics[width=0.95\columnwidth]{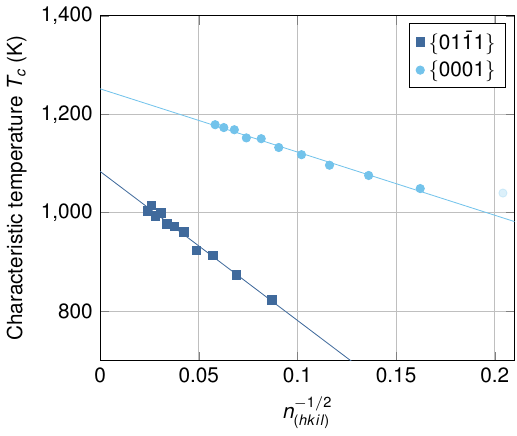}
    \caption{Linear regression of the characteristic temperatures of 
    $\{01\bar{1}1\}$ (squares) and $\{0001\}$ (circles) facets as a function of $n_{(hkil)}^{-1/2}$, where $n_{(hkil)}$ is the number of atoms in the corresponding $(hkil)$ facet. Characteristic temperatures of the smallest nanoparticle (shaded here) was excluded from the linear fit for the melting of $\{0001\}$ facets. Temperatures corresponding to $n_{(hkil)}\rightarrow +\infty$ are 1086 K and 1252 K for $\{01\bar{1}1\}$ and $\{0001\}$ facets, respectively.}
    \label{fgr:GT2D}
\end{figure}

We find that the characteristic temperature $T_c$ scales with $n^{-1/2}$, which we demonstrate to emerge from the Gibbs-Thomson effect in the next section.
The corresponding linear regressions are presented in Fig.~\ref{fgr:GT2D}. 
For infinite surfaces, this approach predicts melting temperatures $T_{c,\infty}$ of 1086 and 1252 K for \((01\bar{1}1)\) and $(0001)$ surfaces, respectively. However, it is important to note that facet melting in nanoparticles is induced by edges and vertices. Consequently, the melting temperature of a perfect infinite surface should be higher than these values.

\subsection{Discussion}

Here, we extend the discussion on the Gibbs-Thomson effect and its relation to premelting.
The derivation of the Gibbs Thomson relation proposed for spherical particles (Eq.~\ref{eqn:eq2}) can be reproduced for 2D surfaces with finite boundaries.

Let us consider a disk of perimeter $L$, representing a facet melting from its edges. 
The outer liquid layer of thickness $r_l -r_s$ near the boundary coexists with the solid facet core of radius $r_s$. 
In 2D, the chemical potential for phase $i$ writes
\begin{equation}
    \mu_i = \mu^{s}_i + \Gamma_i \left(\frac{\partial L_i}{\partial n_i}\right)_{T,P}
\end{equation}
where $\mu^s_i$ is the surface chemical potential, and the second term accounts for the border excess energy, with $\Gamma_i$ the line tension.
Again expressing the Gibbs energy change, assuming equilibrium and similar densities (i.e. $\partial L_l /\partial n_l\approx 0$), we obtain an expression for the melting temperature of the $(hkil)$ facet
\begin{equation}
    T_{c, (hkil)} = T_{c,(hkil)\infty} - \frac{\Gamma_s}{\Delta^{s\rightarrow l} S_m^{(hkil)}}   \left(\frac{\partial L_s}{\partial n_{(hkil)}}\right)_{T,P},
    \label{eq:2dgibbs}
\end{equation}
where $T_{c,(hkil)}$ is the characteristic melting temperature of the $\{hkil\}$ facet containing $n_{(hkil)}$ atoms, $T_{c,(hkil)\infty}$ is the characteristic temperature of the infinite surface with Miller indices $(hkil)$, and $\Delta^{s\rightarrow l} S_m^{(hkil)}$ is the molar surface melting entropy, again assumed constant.
The perimeter is $L_s= 2 \pi r_s$, and the number of atoms contained in the disk is $n_{(hkil)} = \pi r_s^2/a_m$, with $a_m$ the effective atomic area. We obtain $r_s = (n_{(hkil)}a_m/\pi)^{1/2}$.
Then:
\begin{equation}
    \left(\frac{\partial L_s}{\partial r_s}\right) \left(\frac{\partial r_s}{\partial n_{(hkil)}}\right) = 
    2\pi \frac{a_m}{2\pi r_s} = 
    \frac{a_m}{r_s} \propto n_{(hkil)}^{-1/2}.
    \label{eq:circle}
\end{equation}

In the two-dimensional case, we obtain the Gibbs-Thomson relation
\begin{equation}
    T_{c,(hkil)} = T_{c,(hkil)\infty} - b\,{n_{(hkil)}^{-1/2}},
    \label{eqn:TC2D}
\end{equation}
which closely matches our results presented in Fig.~\ref{fgr:GT2D}, except for \{0001\} facets containing less than 45 atoms (i.e. $n^{-1/2}_{(hkil)}\gtrapprox 0.15$), where the geometric assumptions of our simple model seem inaccurate, and statistical error can be larger.

We now relate our surface and facet melting expressions, which both describe the same underlying Gibbs-Thomson effect.
Dividing Eq.~\ref{eqn:TC2D} by Eq.~\ref{eqn:eq2} gives
\begin{equation}
    \dfrac{T_{c,(hkil)}}{T_{m,\mathrm{NP}}}
    = \dfrac{T_{c,(hkil)\infty} + b'x}{T_{m,\infty} + a'x}
\end{equation}
where $x=\dfrac{1}{R}$ and $a'$ and $b'$ are constants.
The Taylor expansion around $x=0$ gives
\begin{equation}
    \dfrac{T_{c,(hkil)}}{T_{m,\mathrm{NP}}}
    = \dfrac{T_{c,(hkil)\infty}}{T_{m,\infty}} 
    + \dfrac{d}{R}
    + \mathcal{O} \left(R^{-2}\right),
\label{eqn:ratio}
\end{equation}
where $d = \dfrac{b' T_{m,\infty} - a'T_{c,(hkil)\infty}}
                 {T_{m,\infty}^2}$ is a constant.
                 
\begin{figure}[b]
    \centering
    \includegraphics[width=0.95\columnwidth]{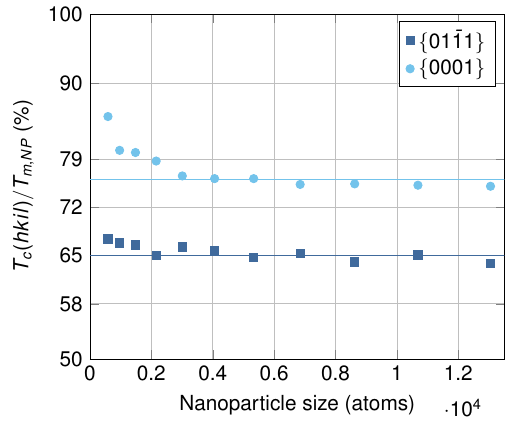}
    \caption{Characteristic ratio of surface-to-NP melting temperature for $\{01\bar{1}1\}$ (squares) and $\{0001\}$ (circles) facets. Average values for $\{01\bar{1}1\}$ and $\{0001\}$ facets of hcp nanoparticles containing more than 1000 atoms are 65~\% and 76~\% respectively.}
    \label{fgr:Tmsurf_on_Tmbulk}
\end{figure}

We investigated the ratio \(T_{c(hkil)} / T_{m,\infty}\) for each facet and plotted it in Fig.~\ref{fgr:Tmsurf_on_Tmbulk} as a function of the NP size.
As expected from Eq.~\ref{eqn:ratio}, the ratio \(T_{c(hkil)} / T_{m,NP}\) becomes independent of the NP size when $R$ is large. Reading  Figs.~\ref{fgr:GT3D} and \ref{fgr:GT2D} for each facet at a larger NP size, we found $\dfrac{T_{c,(01\bar{1}1)\infty}}{T_{m,NP}} = 0.60$ and $\dfrac{T_{c,(0001)\infty}}{T_{m,NP}} = 0.77$, respectively. These values closely match the average trends shown in Fig.~\ref{fgr:Tmsurf_on_Tmbulk}, reinforcing the relevance and validity of adapting the Gibbs-Thomson relation to two-dimensional facets.

Thus, for sufficiently large nanoparticles, the constant \(T_{c(hkil)} / T_{m,NP}\) ratio indicates that nanoparticle surfaces consistently reach their characteristic temperature \(T_{c(hkil)}\) at a fixed fraction of their melting temperature. The fact that $\{01\bar{1}1\}$ facets melt before $\{0001\}$ is consistent with energetics, as the surface energy of the $\{0001\}$ facet is lower than that of $\{01\bar{1}1\}$ (2.13 J.m\textsuperscript{-2} compared to 2.38 J.m\textsuperscript{-2}). This staged surface melting behavior is very similar to the one observed for the roughening transition of metallic (macroscopic) crystal surfaces \cite{lapu1994}.

The model of Couchman and Jesser \cite{Couchman-Jesser} can be thought of as a wetting of the solid core by a liquid shell at $T < T_{m,\infty}$ in a metastable state. The authors found that wetting is limited by the finite size of the nanoparticle, and that the critical liquid layer thickness is not constant but monotonically increasing with particle size.
This has recently been observed experimentally for Sn nanoparticles \cite{sn-np-melting}, and is similar to the behavior modeled in the context of surface segregation in bimetallic nanoparticles \cite{Creuze2012}.

The present simulations reveal two regimes as shown in Fig. 6, namely a linear growth of the thickness of the liquid layer up to NPs containing approximately 5000 atoms and subsequently a smooth convergence towards a thickness of about three atomic layers for larger NPs. The linear regime is consistent with and refines the model of Couchman and Jesser \cite{Couchman-Jesser}, which is applicable to small NPs.
For larger NPs, the present simulations reveal convergence to a constant thickness. The limits for each of the facets would be the corresponding surfaces of macroscopic crystals. 

It is remarkable that electronic structure calculations of metal surfaces show that the local density of electronic states converges  to bulk like behavior within about three atomic layers for transition metals such as Ni, while the convergence is slower for simple metals such as Al, thus connecting and reconciling the present simulations of NPs with our understanding of metal surfaces \cite{wimmer2000fundamentals}.

Finally, we note an advantage of the present algorithm to describe the nanoparticle melting with far greater precision than traditional approaches such as a-CNA \cite{a-cna}, PTM \cite{ptm}, or the Lindemann criterion \cite{lindemann}.
The robustness of machine-learning-based analysis methods against thermal noise is now well established \cite{robust, distortion-score, tom, neighbor-map}. However, the materials studied previously often possessed a majoritary distribution (e.g., defects in bulk) and a reduced diversity of atomic environments. Furthermore, when using the Mahalanobis distance, a key assumption is that the statistical distributions of atomic environments are unimodal Gaussians \cite{distortion-score}, which is not rigorously verified in the present case. The lack of periodicity and the small sizes of the systems studied thus made their analysis particularly challenging.

In the classification used in this paper, all atoms of a given facet type are grouped in the same class. For instance, an atom located next to the edge of a $\{01\bar{1}1\}$ facet is assigned to the $\{01\bar{1}1\}$ facet's class, as well as an atom located in the center of the same facet. Further decomposing the $\{01\bar{1}1\}$ facet's class into several classes using the GMM algorithm could reveal the presence of several sub-distributions, allowing to separate atoms to reflect differences on their local environment.
This suggests that a finer granularity of analysis would be straightforward to achieve using the same workflow, allowing the analysis of local atomic patterns found on surfaces.
In the present work, we avoided excessively multiplying the number of classes, and chose to work with classes that do not rigorously adhere to the unimodality condition. 
The fact that our classification performs well, even for non-unimodal classes, further underscores the robustness and flexibility of these methods, two characteristics that are clear limitations of more traditional analysis techniques such as PTM~\cite{ptm} or a-CNA~\cite{a-cna}.

The present work relies on a combination of MD simulations and advanced post-processing techniques to investigate the melting of metallic nanoparticles in fine detail, leading to several general conclusions.
We took several steps to ensure the statistical robustness of the results, although their translation into error bars or formal error quantification remains challenging.
To assess the effect of the heating procedure, we verified that we found no significant variations in observations when maintaining the system for more than 50 ps at each temperature, and used a more conservative value of 100 ps per 10K increment.

To provide a quantitative indication of the statistical dispersion of the results obtained, we generated 8 MD trajectories for the 2157-atom nanoparticle and the analysis of these trajectories is presented in Table S2. As can be observed, the standard deviations for both the nanoparticle global melting temperature and the facet critical temperatures are on the order of the temperature step used for the MD trajectories, which is the best outcome that could be expected. These results show also minor variations of the critical effective liquid layer thickness. Reproducing the same approach for all sizes would have required substantial resources to certainly lead to similar observations.

For structure recognition, we used the distortion score, a robust metric for local atomic environment similarity, where out-of-distribution samples are directly associated with a large statistical distance.
To ensure reliability, all multivariate Gaussians were fitted on a statistically significant count of samples, even for the smallest class, i.e., nanoparticle vertices, which contains $n=126,000$ samples.
To mitigate the presence of any out-of-distribution samples in training data, we minimized the empirical covariance determinant by systematically ignoring a fraction of samples set to 7\% \cite{distortion-score}.

A potential source of error is the choice of a threshold to define a partition between in-class atoms and outliers. To adhere to an unsupervised approach, we defined this threshold in a systematic way based on the last local minimum before the second peak in Mahalanobis distance distribution, limiting the risk of false positives, i.e. atoms being wrongly classified as outliers. However, for distributions characterized by distant peaks, such as the atoms of the nanoparticle core and $\{0001\}$ facets, a flat region in the distance distribution makes this minimum ill-defined. In such cases, the threshold distance was left as an adjustable parameter (see Methods and Fig.~\ref{fgr:dist_score}).
While we showed that the present approach produced robust predictions, applications on different systems might motivate the use of alternative strategies for label assignment \cite{poisvert2025short}.

\subsection{Conclusion}

This study quantifies the surface premelting of hexagonal close-packed cobalt nanoparticles containing between 587 and 13047 atoms, confirming recent experimental results~\cite{sn-np-melting}, and further highlighting the role of crystallographic facet orientation in these processes. Using a machine-learned classifier, we demonstrate that stepped $\{01\bar{1}1\}$ facets consistently melt before flat $\{0001\}$ facets, regardless of nanoparticle size. The melting temperature gap between these facets remains invariant with size, while the global melting behavior follows the classical Gibbs--Thomson law.

Three key findings emerge from this work: (i) atoms at vertices begin diffusing well below the bulk melting temperature, near 400~K, approaching ambient conditions; (ii) a critical fraction of structurally disordered atoms must be reached to trigger complete melting of the nanoparticle and this fraction varies non-monotonically with size, reflecting a critical liquid layer thickness that increases monotonically with nanoparticle size converging to a thickness of approximately three atomic layers; and (iii) for very small nanoparticles, global melting requires the complete disordering of the outermost shell, including the most stable facets.

We further extend the Gibbs--Thomson relation to planar surfaces, linking facet-specific melting temperatures to nanoparticle size. Compared to traditional structural analysis techniques, our descriptor-based approach offers greater robustness against thermal noise and versatility to analyze complex atomic environments. These results demonstrate the ability of such methods to characterize systems with diverse atomic environments, opening new avenues for the study of nanoparticles properties at elevated temperatures.

\section{Methods}\label{sec:methods}

\begin{figure}[b]
    \includegraphics[width=0.25\linewidth]{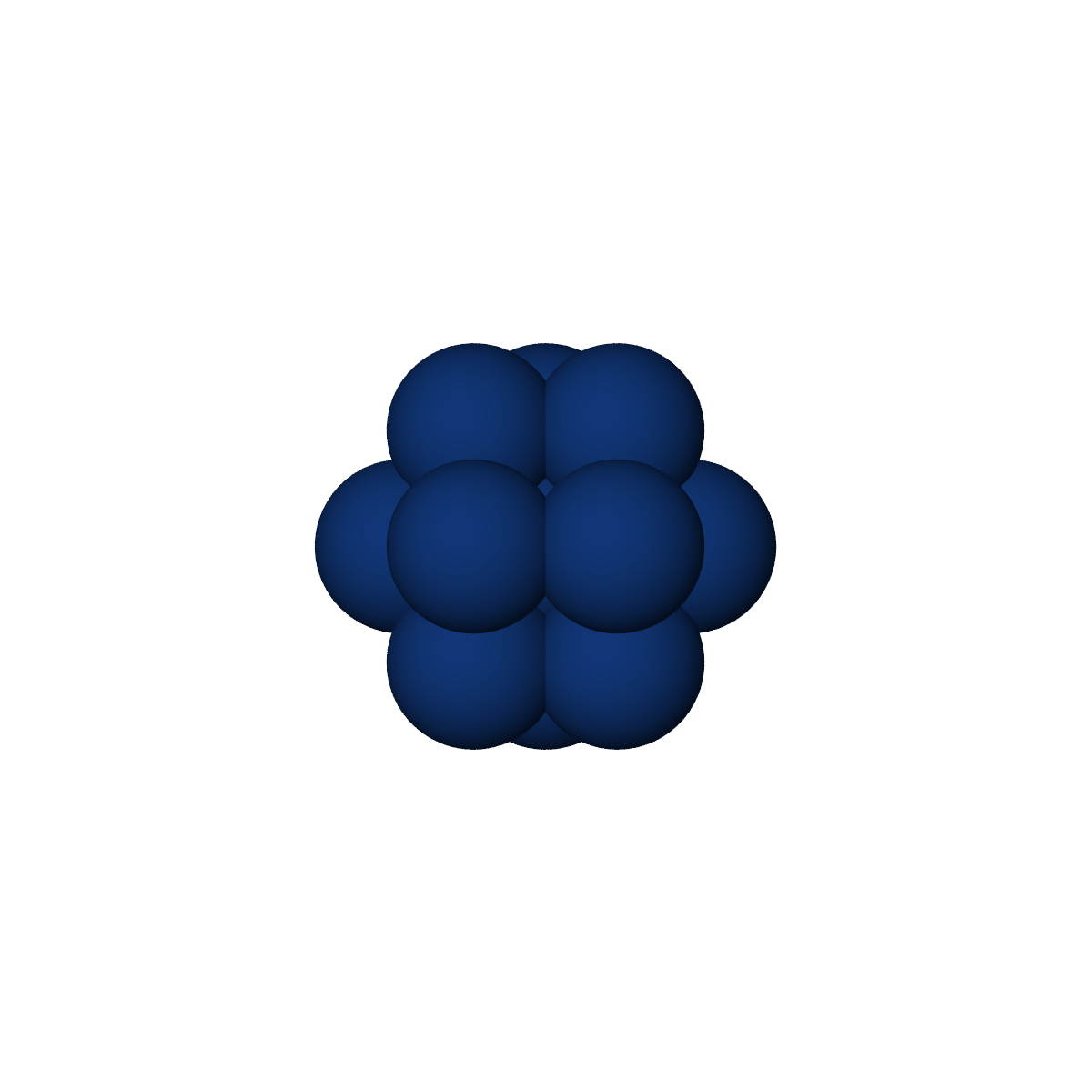}
    \includegraphics[width=0.25\linewidth]{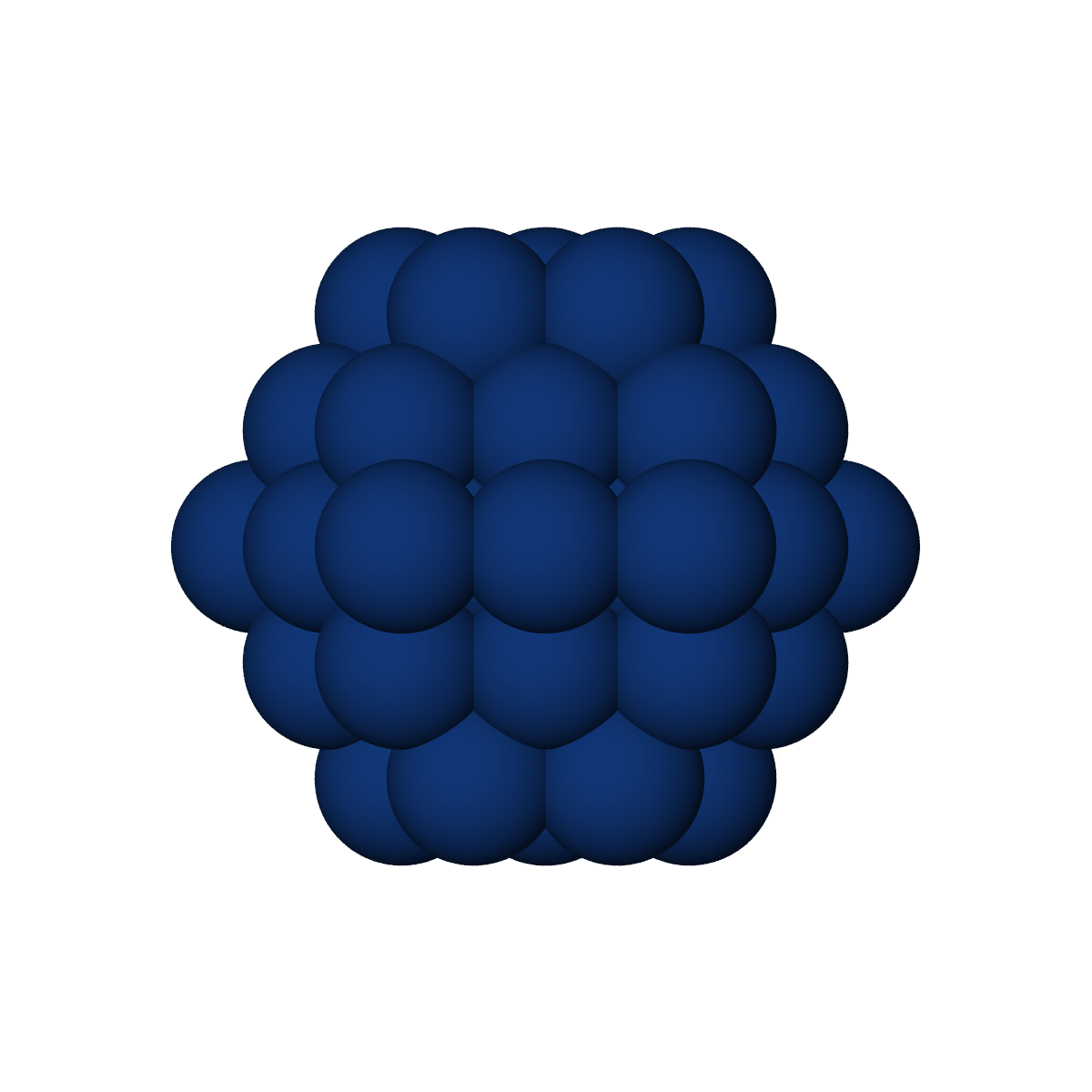}
    \hspace{0.075\linewidth}
    \includegraphics[width=0.25\linewidth]{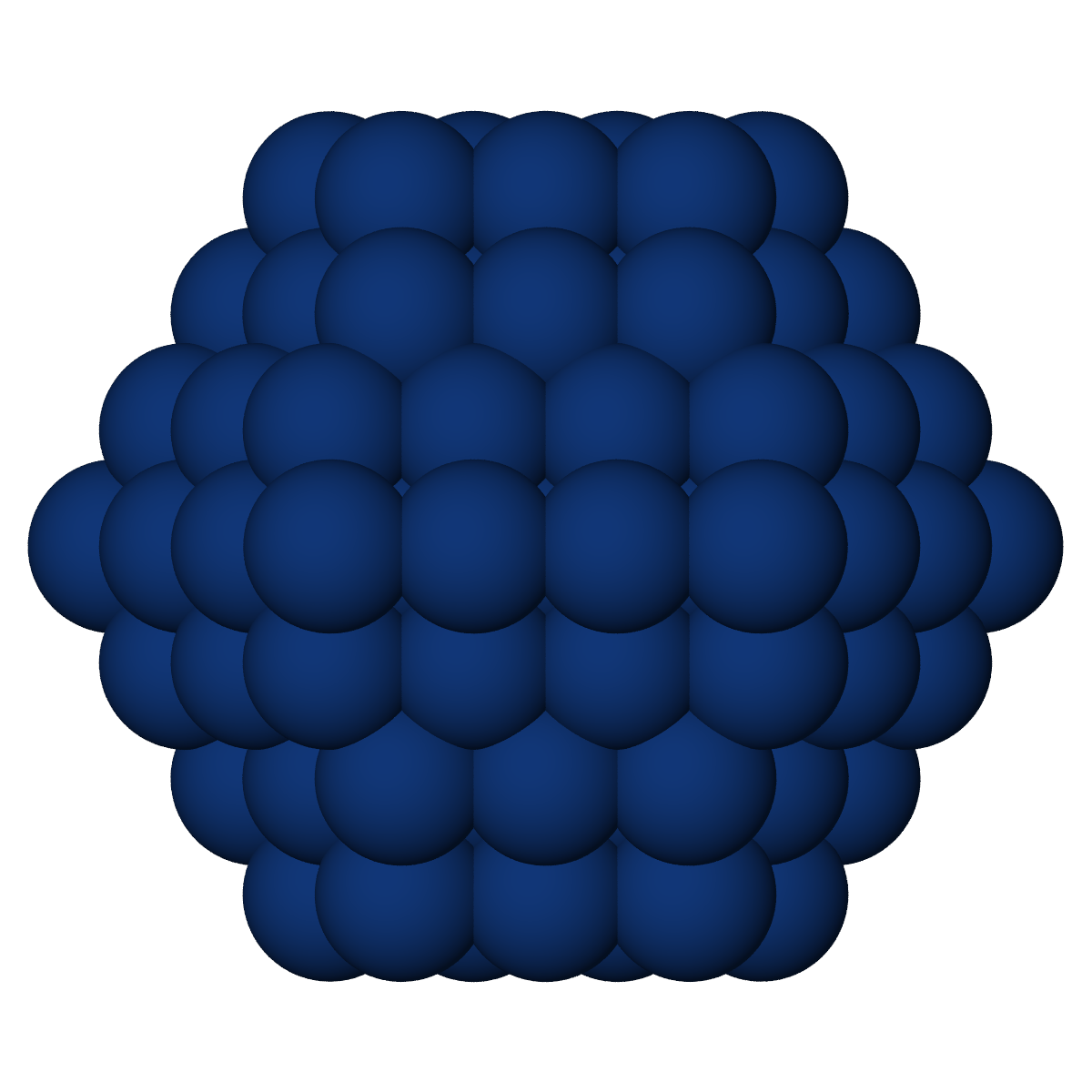}
    \caption{Construction of hexagonal close-packed (hcp) nanoparticles using the \textit{MedeA} materials modeling environment \cite{medea}. The process begins by selecting the smallest hexagonal motif in an hcp supercell, consisting of 13 atoms (left). Successive layers of nearest neighbors are then added to grow the nanoparticle to the desired size. The middle and right nanoparticles illustrated in this figure contain 57 and 153 atoms, respectively.}
    \label{fig:construction_NPhcp}
\end{figure}

\paragraph{Creation of nanoparticles}
Hexagonal close-packed nanoparticles were created using the \textit{MedeA} materials modeling environment \cite{medea}, following the protocol illustrated on Fig.~\ref{fig:construction_NPhcp}. First, the smallest hexagon is selected inside a large hcp supercell. Then, the nearest unselected neighbor of each atom is added to the hexagon: a layer is added. This operation is repeated until the desired size is obtained. 

\paragraph{Molecular dynamics simulations}
The heating trajectory is simulated using the LAMMPS molecular dynamics package \cite{lammps} and a quadratic spectral neighbor analysis potential (q-SNAP) \cite{snap, qsnap} developed for bulk, surface, and nanoparticle structures of cobalt \cite{CoP}. The generation of the machine-learned potential and its use with LAMMPS are integrated functionalities of \textit{MedeA}. The heating is performed in steps of 100 ps per 10 K increment, ranging from 50 K to 1600 K. A Nosé-Hoover thermostat with a damping constant of 200 fs is employed, and the time step is set to 2 fs. The bispectrum descriptors of all atoms are saved every 2 ps. They are computed using an angular moment parameter $j_{max} = 4$ and a cutoff radius $R_{cut} = 5$~\AA, as used in the q-SNAP potential \cite{CoP} used here.

\paragraph{Structural analysis}
The bispectrum descriptors are used to train the classifier at low temperatures. The purpose of this initial step is to define the classes of atoms (e.g., bulk, facets, edges, and vertices). To ensure accurate classification, the training temperature must be low enough so that surface diffusion has not yet started, as diffusing atoms would no longer correspond to their original classes. For hcp nanoparticles, this temperature is set to \(T_{\text{max,hcp}} =400\) K, at which vertex atoms begin to diffuse towards the edges and the $\{01\bar{1}1\}$ facets. The remainder of the heating trajectory is used solely for classification.

We use unsupervised learning with Gaussian Mixture Models (GMM) to define the atomic classes in a hierarchical, binary manner. At each step, the training set is divided into two classes using two gaussians, even if the final number of classes is larger. For example, we first separate the atoms into bulk and surface. Then, one of these subsets, such as the surface atoms, is further split into two classes, and this process is repeated iteratively until all atomic classes (e.g., facets, edges, vertices) are identified.
When applied to other systems, it could become difficult to intuitively determine how many iterations  should be performed, which can be addressed by applying standard statistical methods for model evaluation.

\begin{figure}[hbtp]
    \centering
    \includegraphics[width=0.95\columnwidth]{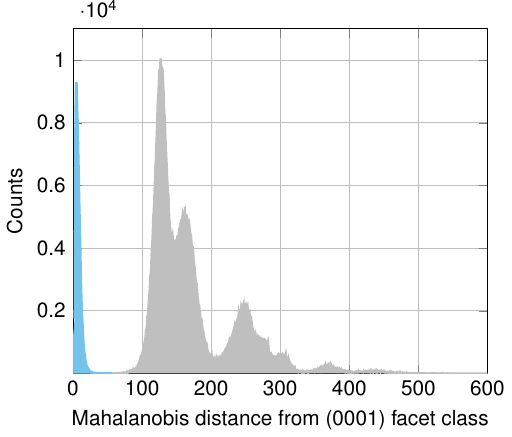}
    \caption{Distorsion scores of atoms in the training set relative to the $\{0001\}$ facet class for a 4061 atoms hcp nanoparticle. The first peak, displayed in light blue, corresponds to the atoms from $\{0001\}$ facets, whereas following gray peaks correspond to atoms from other classes.}
    \label{fgr:dist_score}
\end{figure}

The Mahalanobis distance $d_{k}(\mathbf{x})$ of $\mathbf{x}$ with respect to the class $k$ with empirical mean $\boldsymbol{\mu}_k$ and covariance $\boldsymbol{\Sigma}_k$ is expressed as:
\begin{equation}
d_{k}(\mathbf{x}) = \sqrt{(\mathbf{x}-\boldsymbol{\mu}_k)^{\top} \boldsymbol{\Sigma}^{-1}_k (\mathbf{x}-\boldsymbol{\mu}_k)}.
\label{eq:maha_dist}
\end{equation}

Once the classes are defined, we first use the Minimum Covariance Determinant (MCD) algorithm to refine the estimates of the empirical mean and covariance matrix, which are critical parameters for Mahalanobis distances calculations, using a contamination level set to 7~\% of the data \cite{distortion-score, mcd}.
Then, we compute the empirical covariance $\boldsymbol{\Sigma}$ of each class. 
Note that we ensure that the sample count is large enough for all distributions, to ensure the statistical robustness of $\boldsymbol{\Sigma}$. The class with the lowest sample count, i.e., nanoparticle vertices, contains $126,000$ examples.
It allows us to determine a threshold Mahalanobis distance (or distortion score) \cite{distortion-score} for all atoms in the training set relative to each class. 
As shown in Fig.~\ref{fgr:dist_score}, Mahalanobis distance histograms exhibit a primary peak at short distances, corresponding to atoms that belong to the class associated with the histogram. 
Subsequent peaks represent atoms that do not belong to the current class. The threshold distance is automatically determined as the closest minimum to the second peak, except for the $\{0001\}$ facets and for the nanoparticle core, which do not exhibit a well-defined minimum in the distance distribution. 
Instead, we found that taking a threshold distance equal to 60 and between 55 and 58 for the bulk and $\{0001\}$ facets, respectively, led to predictions  that were physically-consistent and robust to thermal noise.
This criterion was selected to minimize the rate of false positives, i.e. atoms being wrongly classified as outliers, resulting in improved stability at high temperature.

The Mahalanobis distance for each atom is then computed across the entire trajectory, and atoms are classified based on these threshold distances. This classification approach is more robust than simpler binary classifiers, which always assign a class to an atom. Here, if an atom’s distance exceeds all the maximum threshold distances, it is identified as an outlier. Such a classification would not be possible if we relied solely on the GMM model, for example.

\subsection*{Supporting Information}
Fig.~\ref{fgr:doutliers} and Fig.~\ref{fgr:surface_melting} are given for all nanoparticles in Supporting Information. Relevant information such as global melting temperatures, critical effective liquid layer thicknesses and critical temperatures of each facets for all nanoparticle sizes are summarized in Table S1. The same observables extracted from 8 different MD trajectories for the 2157-atom nanoparticle are summarized in Table S2. Visualizations of the liquid layer for all nanoparticle sizes are shown in Fig.~S4.

\subsection*{Acknowledgments}
We gratefully acknowledge Mihai-Cosmin Marinica for his guidance in adapting the distortion score method to this study. The authors are grateful to all colleagues at Materials Design. AA acknowledges fruitful discussions with Fabienne Ribeiro and Andrei Jelea.
The French Association Nationale de la Recherche et de la Technologie (ANRT) is acknowledged for CIFRE funding No. 2022/1599.
\bibliographystyle{apsrev4-2}
\bibliography{biblio}
\end{document}

\title{\Large{Supporting information}\\ 
\LARGE{Unveiling and quantifying the topology-dependent premelting of nanoparticles}}

   \author{Marthe Bideault}
    \affiliation{%
    Materials Design SARL, 42 avenue Verdier, 92120 Montrouge, France%
    }%
    \affiliation{
    ICMMO, Université Paris-Saclay, UMR 8182, 17 avenue des Sciences, 91400 Orsay, France
    }%
    \author{Arnaud Allera}
    \affiliation{%
    ASNR/PSN-RES/SEMIA/LSMA Centre d’études de Cadarache, F-13115 Saint Paul-lez-Durance, France
    }
    \email{arnaud.allera@asnr.fr}
    \author{Ryoji Asahi}
    \affiliation{%
    Institute of Materials Innovation, Nagoya University, Nagoya 464-8603, Japan
    }%
    \author{Jérôme Creuze}
    \affiliation{%
    ICMMO, Université Paris-Saclay, UMR 8182, 17 avenue des Sciences, 91400 Orsay, France
    }%
    \email{jerome.creuze@universite-paris-saclay.fr}
    \author{Erich Wimmer}
    \affiliation{%
    Materials Design SARL, 42 avenue Verdier, 92120 Montrouge, France
    }%
    \email{ewimmer@materialsdesign.com}
    \affiliation{%
    Materials Design, Inc., 12121 Scripps Summit Drive, \#160, San Diego, California 92131, USA }%

\maketitle

\begin{figure}
    \centering
    \includegraphics[width=.85\linewidth]{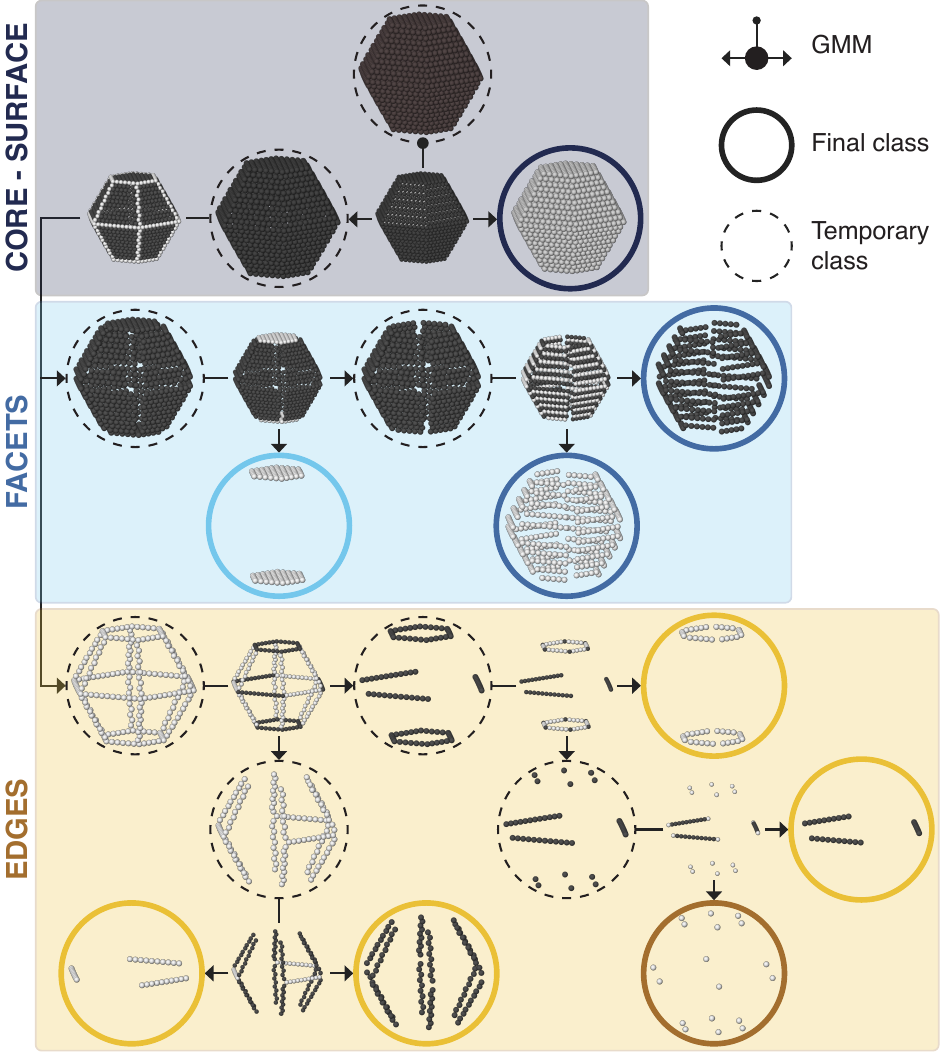}
    \caption{Classification tree for the initial unsupervised classification using Gaussian Mixture Models (GMM). Starting from an initial nanoparticle containing unlabeled atoms, binary GMM classification is performed. The training set is separated into two subclasses (colored in black and white) by the binary GMM. The process is repeated until the classes are considered as final. Final classes corresponding to bulk, $\{0001\}$ facets, $\{01\bar{1}1\}$ facets, edges, and vertices are circled in indigo, light blue, dark blue, yellow and brown, respectively. Temporary classes (including atoms corresponding to different final classes) are circled with dashed lines.}
    \label{fig:classification_tree}
\end{figure}
\begin{figure}
    \centering
    \begin{subfigure}[b]{0.3\linewidth}
        \includegraphics[height=4.cm]{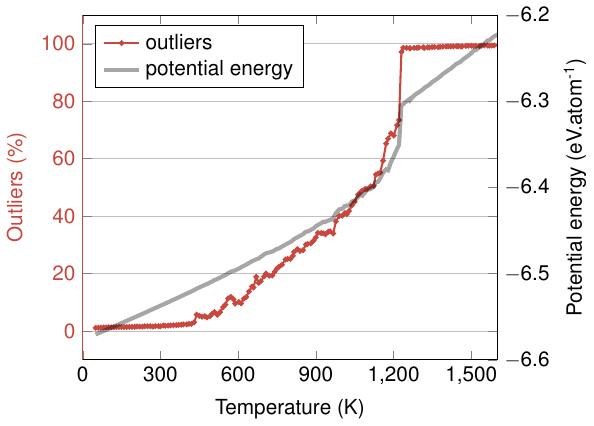}
        \caption{587 atoms}
    \end{subfigure}
    \hspace{0.03\linewidth}
    \begin{subfigure}[b]{0.3\linewidth}
        \includegraphics[height=4.cm]{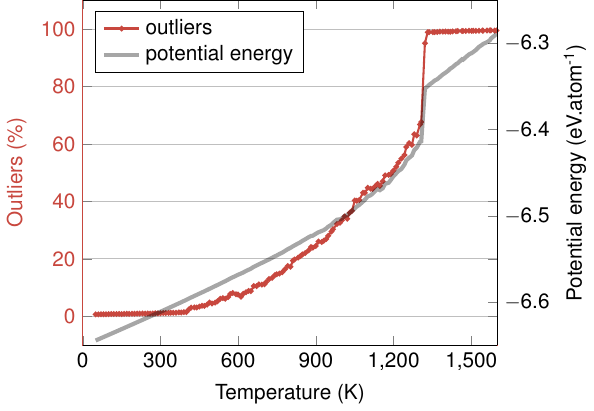}
        \caption{967 atoms}
    \end{subfigure}
    \hspace{0.03\linewidth}
    \begin{subfigure}[b]{0.3\linewidth}
        \includegraphics[height=4.cm]{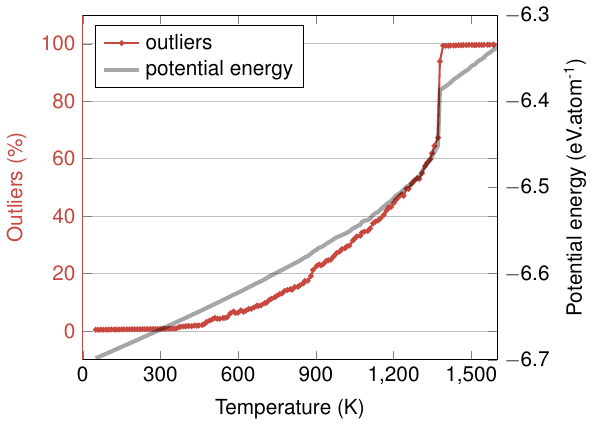}
        \caption{1483 atoms}
    \end{subfigure}
    \vspace{0.8cm}
    
    \begin{subfigure}[b]{0.3\linewidth}
        \includegraphics[height=4cm]{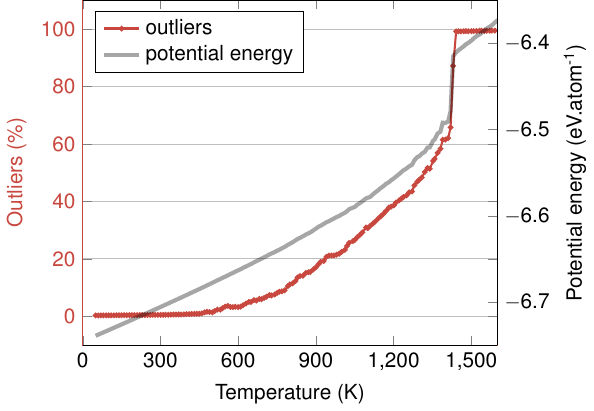}
        \caption{2157 atoms}
    \end{subfigure}
    \hspace{0.03\linewidth}
    \begin{subfigure}[b]{0.3\linewidth}
        \includegraphics[height=4cm]{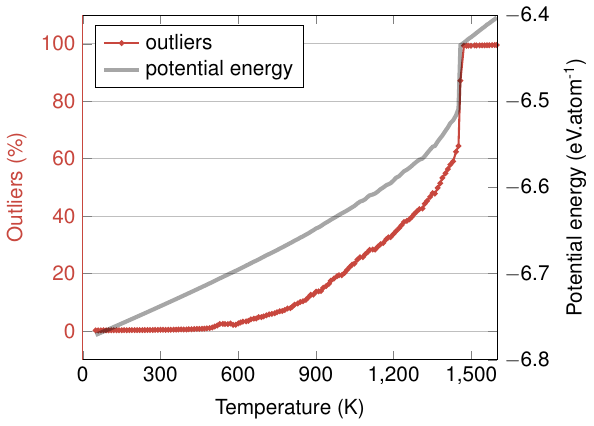}
        \caption{3009 atoms}
    \end{subfigure}
    \hspace{0.03\linewidth}
    \begin{subfigure}[b]{0.3\linewidth}
        \includegraphics[height=4cm]{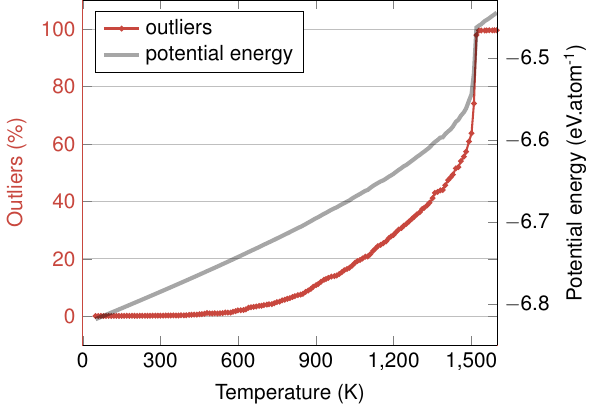}
        \caption{5333 atoms}
    \end{subfigure}
    \vspace{0.8cm}
    
    \begin{subfigure}[b]{0.3\linewidth}
        \includegraphics[height=4cm]{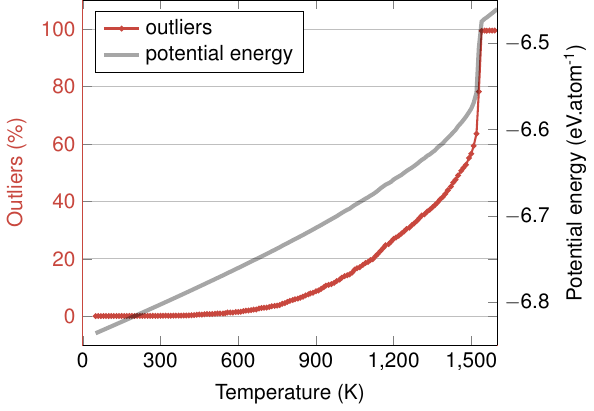}
        \caption{6847 atoms}
    \end{subfigure}
    \hspace{0.03\linewidth}
    \begin{subfigure}[b]{0.3\linewidth}
        \includegraphics[height=4cm]{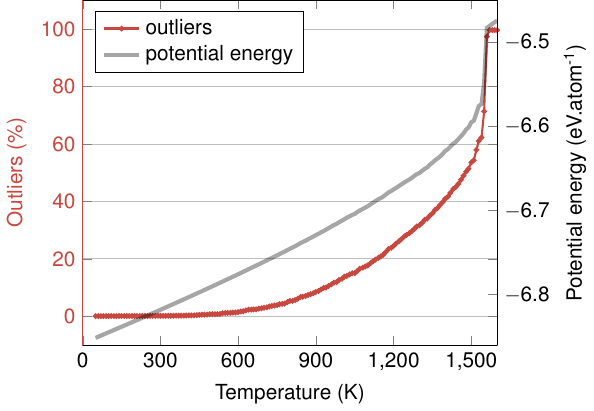}
        \caption{8623 atoms}
    \end{subfigure}
    \hspace{0.03\linewidth}
    \begin{subfigure}[b]{0.3\linewidth}
        \includegraphics[height=4cm]{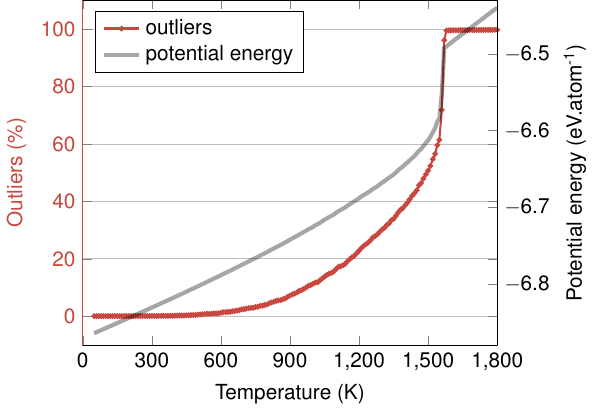}
        \caption{10683 atoms}
    \end{subfigure}
    \vspace{0.8cm}
    
    \begin{subfigure}[b]{0.3\linewidth}
        \includegraphics[height=4cm]{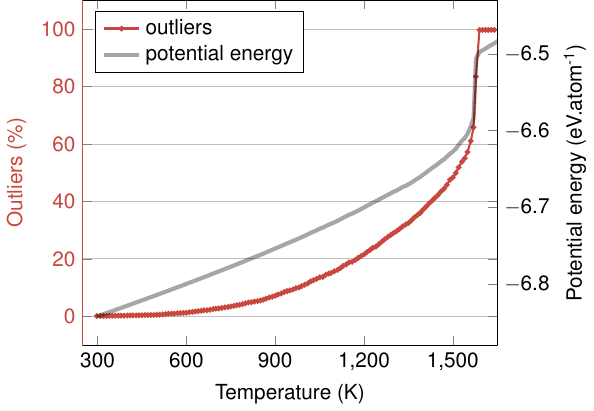}
        \caption{13047 atoms}
    \end{subfigure}
    \caption{Percentage of outliers (plain red line) and potential energy (dashed grey line)
    as a function of temperature, for all sizes of nanoparticles considered in this study. %
    }
\end{figure}

\begin{figure}
    \centering
    \begin{subfigure}[b]{0.3\linewidth}
        \includegraphics[height=4.5cm]{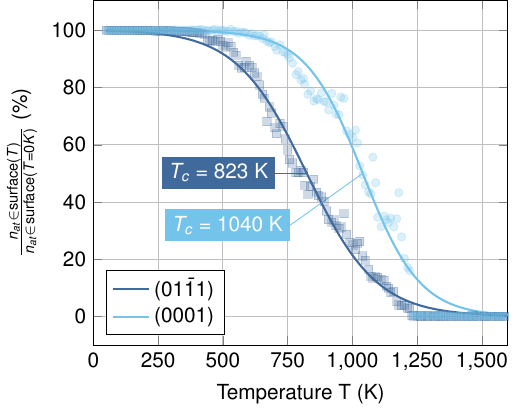}
        \caption{587 atoms}
    \end{subfigure}
    \hspace{0.03\linewidth}
    \begin{subfigure}[b]{0.3\linewidth}
        \includegraphics[height=4.5cm]{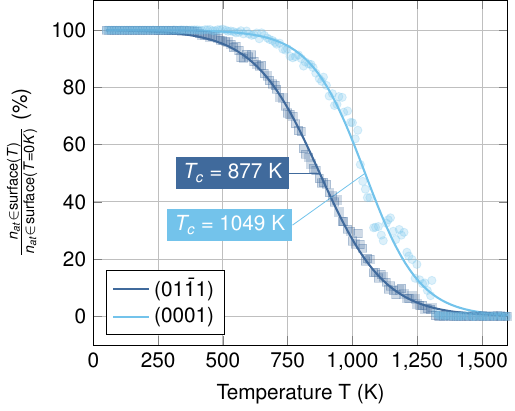}
        \caption{967 atoms}
    \end{subfigure}
    \hspace{0.03\linewidth}
    \begin{subfigure}[b]{0.3\linewidth}
        \includegraphics[height=4.5cm]{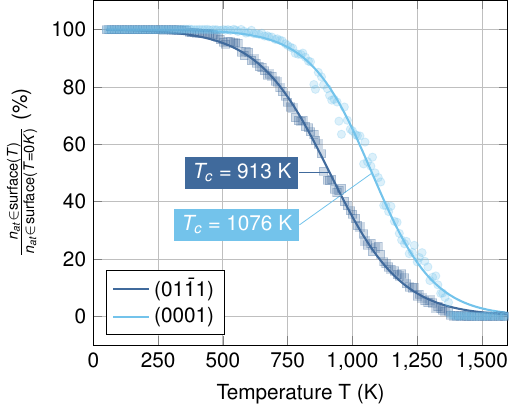}
        \caption{1483 atoms}
    \end{subfigure}
    \vspace{0.8cm}
    
    \begin{subfigure}[b]{0.3\linewidth}
        \includegraphics[height=4.5cm]{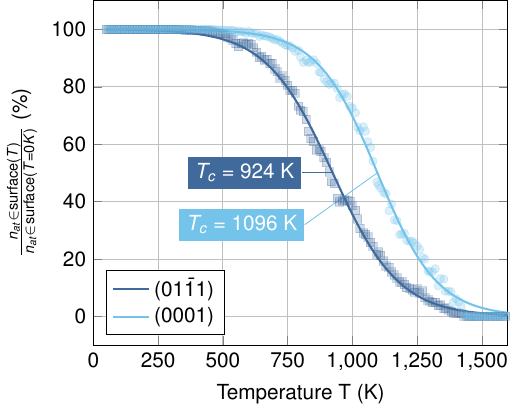}
        \caption{2157 atoms}
    \end{subfigure}
    \hspace{0.03\linewidth}
    \begin{subfigure}[b]{0.3\linewidth}
        \includegraphics[height=4.5cm]{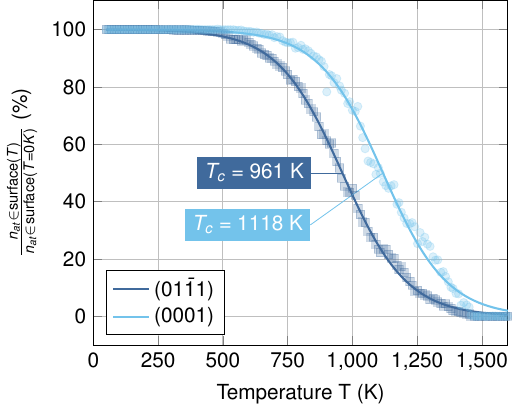}
        \caption{3009 atoms}
    \end{subfigure}
    \hspace{0.03\linewidth}
    \begin{subfigure}[b]{0.3\linewidth}
        \includegraphics[height=4.5cm]{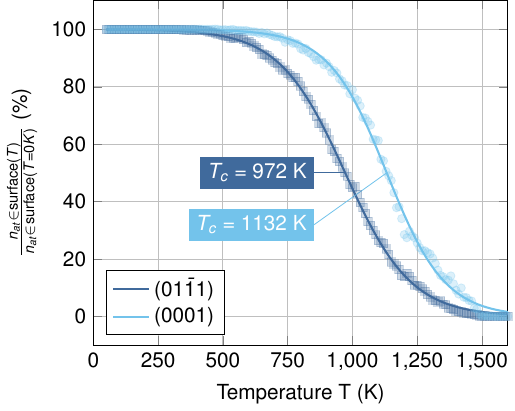}
        \caption{4061 atoms}
    \end{subfigure}
    \vspace{0.8cm}
    
    \begin{subfigure}[b]{0.3\linewidth}
        \includegraphics[height=4.5cm]{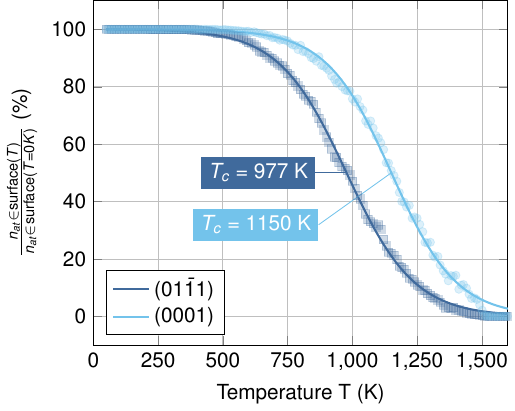}
        \caption{5333 atoms}
    \end{subfigure}
    \hspace{0.03\linewidth}
    \begin{subfigure}[b]{0.3\linewidth}
        \includegraphics[height=4.5cm]{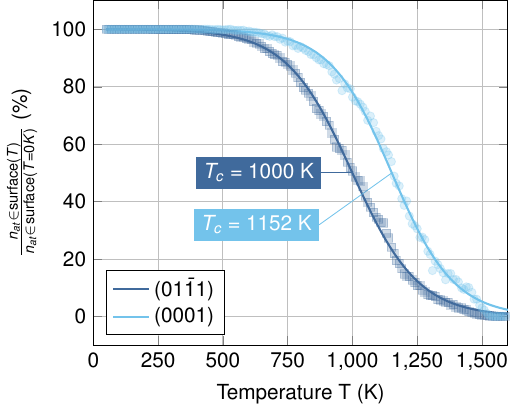}
        \caption{6847 atoms}
    \end{subfigure}
    \hspace{0.03\linewidth}
    \begin{subfigure}[b]{0.3\linewidth}
        \includegraphics[height=4.5cm]{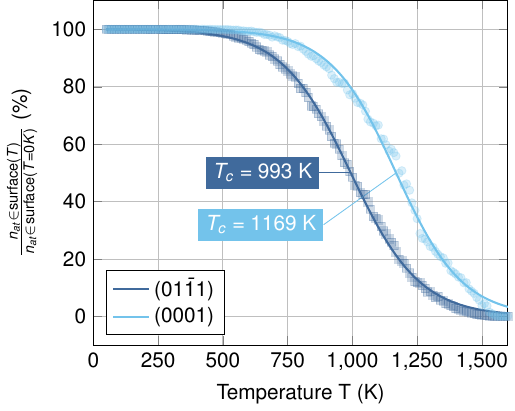}
        \caption{8623 atoms}
    \end{subfigure}
    \vspace{0.8cm}

    \begin{subfigure}[b]{0.3\linewidth}
        \includegraphics[height=4.5cm]{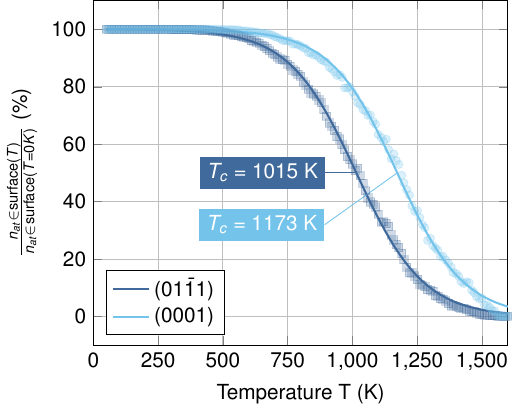}
        \caption{10683 atoms}
    \end{subfigure}
    \hspace{0.03\linewidth}
    \begin{subfigure}[b]{0.3\linewidth}
        \includegraphics[height=4.5cm]{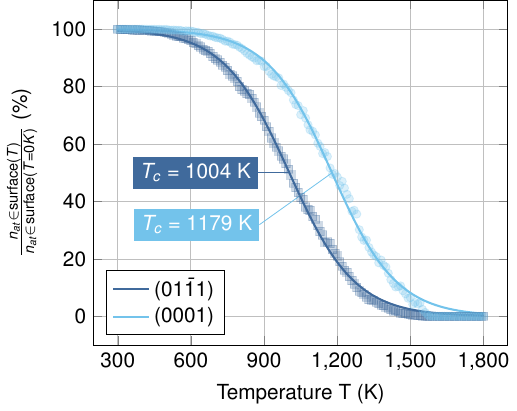}
        \caption{13047 atoms}
    \end{subfigure}
    \caption{Percentage of atoms that still belong to
    $\{01\bar{1}1\}$ (dark blue) and $\{0001\}$ (light blue) facets for all sizes of nanoparticles considered in this study. Plain lines correspond to sigmoid fit of these data. Characteristic temperatures $T_c$ are indicated for each surface.}
\end{figure}

\newpage
\begin{table}
  \caption{Relevant numerical values for each size of nanoparticle.}
    \begin{tabular}{|l||lllllllllll|}
        \hline
        $N$\textsuperscript{\emph{a}} & 587 & 967 & 1483 & 2157 & 3009 & 4061 & 5333 & 6847 & 8623 & 10683 & 13047\\
        \hline
        $n_c$\textsuperscript{\emph{b}} & 323 & 587 & 967 & 1483 & 2157 & 3009 & 4061 & 5333 & 6847 & 8623 & 10683 \\
        \hline
        $n_s$\textsuperscript{\emph{c}} & 264 & 380 & 516 & 674 & 852 & 1052 & 1272 & 1514 & 1776 & 2060 & 2364 \\
        \hline
        $T_{m,NP}$ (K)\textsuperscript{\emph{d}} & 1220 & 1310 & 1370 & 1420 & 1450 & 1480 & 1510 & 1530 & 1550 & 1560 & 1570 \\
        \hline
        $l_c$ ($a_0$)\textsuperscript{\emph{e}} & 1.82 & 1.89 & 2.01 & 2.14 & 2.27 & 2.46 & 2.61 & 2.80 & 2.86 & 2.97 & 3.06 \\
        \hline
        $T_{c,(01\bar{1}1)}$\textsuperscript{\emph{f}} (K) & 823 & 874 & 913 & 924 & 961 & 972 & 977 & 1000 & 993 & 1015 & 1004 \\
        \hline
        $T_{c,(0001)}$\textsuperscript{\emph{f}}(K)& 1040 & 1049 & 1075 & 1096 & 1117 & 1132 & 1150 & 1152 & 1169 & 1173 & 1179 \\
        \hline

    \end{tabular}
    
  \textsuperscript{\emph{a}} Number of atoms;
  \textsuperscript{\emph{b}} Number of core atoms;
  \textsuperscript{\emph{c}} Number of surface atoms;
  \textsuperscript{\emph{d}} Nanoparticle's melting point;
  \textsuperscript{\emph{e}} Critical effective thickness of the liquid layer before global melting of the nanoparticle, expressed in units of the nearest-neighbour distance in hcp Co at $T=0$ K ($a_0=2.49$ \AA);
  \textsuperscript{\emph{f}} Characteristic temperature for $(01\bar{1}1)$ and $(0001)$ facets.
\end{table}

\begin{table}
  \caption{Relevant numerical values for 8 different MD trajectories for the 2157-atom nanoparticle. The two last columns represent the average and the standard deviation values, respectively.}
    \begin{tabular}{|l||llllllll|ll|}
        \hline
        $T_{m,NP}$ (K)\textsuperscript{\emph{a}} & 1420 & 1431 & 1430 & 1418 & 1421 & 1410 & 1409 & 1409 & 1419 & 9\\
        \hline
        $l_c$ ($a_0$)\textsuperscript{\emph{b}} & 2.14 & 2.14 & 2.31 & 2.20 & 2.21 & 2.07 & 1.98 & 1.92 & 2.13 & 0.14 \\
        \hline
        $T_{c,(01\bar{1}1)}$\textsuperscript{\emph{c}} & 924 & 936 & 939 & 930 & 930 & 924 & 943 & 936 & 933 & 7\\
        \hline
        $T_{c,(0001)}$\textsuperscript{\emph{c}} & 1096 & 1095 & 1118 & 1100 & 1111 & 1119 & 1117 & 1102 & 1107 & 10\\
        \hline
    \end{tabular}
    
  \textsuperscript{\emph{a}} Nanoparticle's melting point;
  \textsuperscript{\emph{b}} Critical effective thickness of the liquid layer before global melting of the nanoparticle, expressed in units of the nearest-neighbour distance in hcp Co at $T=0$ K ($a_0=2.49$ \AA);
  \textsuperscript{\emph{c}} Characteristic temperature for $(01\bar{1}1)$ and $(0001)$ facets.
\end{table}

\begin{figure}
    \centering
    \includegraphics[width=0.7\linewidth]{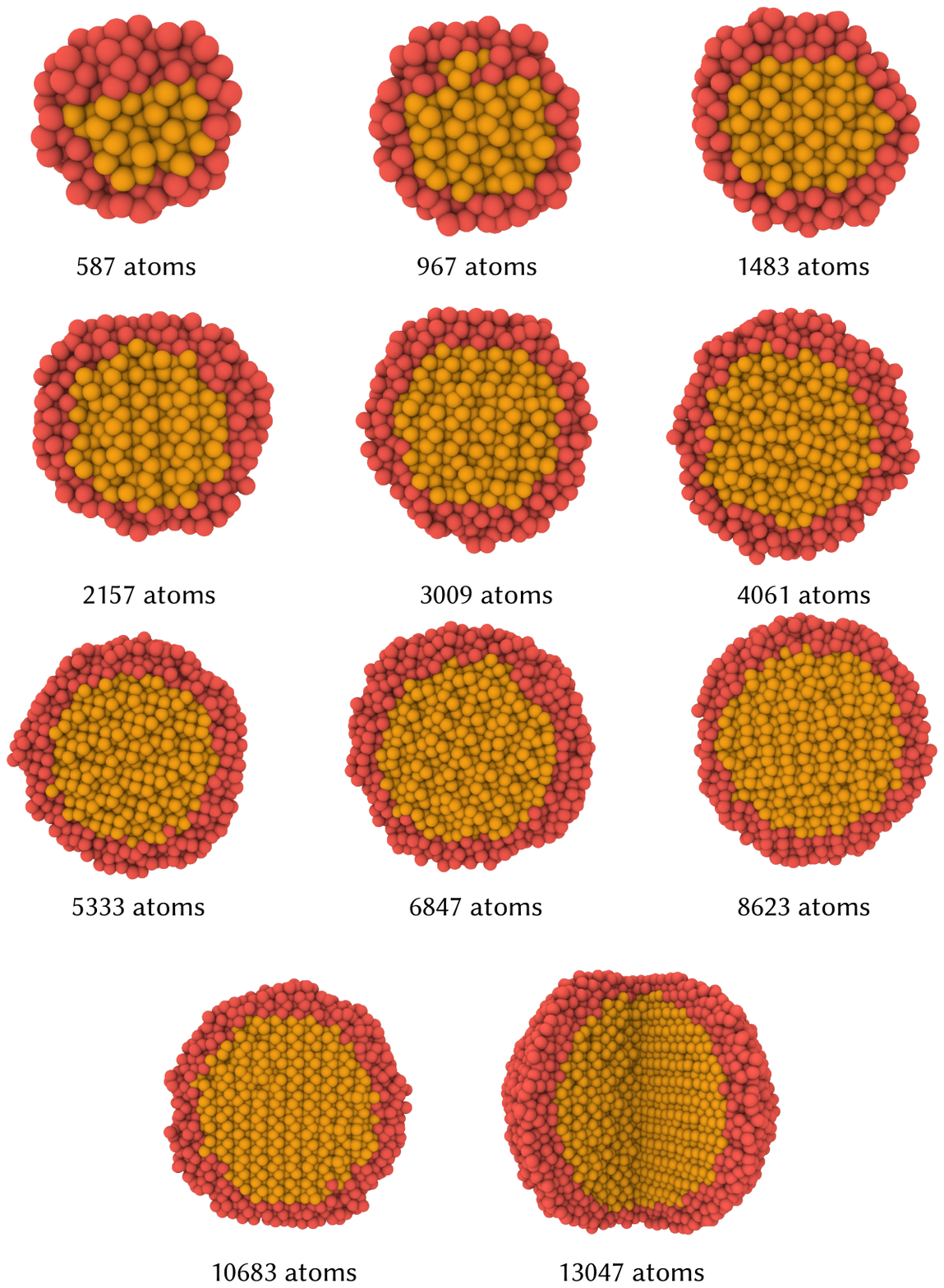}
    \caption{\textbf{Visualizing the liquid layer thickness.} OVITO \cite{ovito} visualizations of nanoparticles at the last temperature before fully melting at $T_{m,NP}$. To enhance stability, labels are defined over the constant temperature trajectory: we represent in red the atoms that are classified as outliers in more than 50\% of the frames. The atomic positions are those of the last frame of the MD trajectory.}
    \label{fig:placeholder}
\end{figure}

\bibliographystyle{apsrev4-2}
\bibliography{biblio}